\documentclass[pra,aps,twocolumn,twoside,english]{revtex4}
\usepackage{amssymb}
\usepackage{amsmath}
\usepackage{amscd}
\usepackage{babel}
\usepackage{latexsym}  % maybe this one is needed, too.
\usepackage{epsfig}
\usepackage{bbm}
\newtheorem{thm}{Theorem}

\newtheorem{lemma}{Lemma}
\newtheorem{cor}{Corollary}
\newtheorem{eg}{Example}
\begin{document}
 %%%%%%%%%%%%%%%%%%%DEFINITIONS%%%%%%%%%%%%%%%%%%%%%%%%%%%%%%%%%
\newcommand{\1} I%{{\openone}}
\def\tr{ {\rm{Tr \,}}}
\def\diag{ {\rm{diag \,}}}
\def\Pr{ {\rm{Pr }}}
\def\supp{ {\rm{supp \,}}}
\def\dim{ {\rm{dim \,}}}
\def\oti{{\otimes}}
\def\bra#1{{\langle #1 |  }}
\def\lb{ \left[ }
\def\rb{ \right]  }
\def\tilde{\widetilde}
\def\bar{\overline}
\def\*{\star}

\def\({\left(}		 		 \def\BL{\Bigr(}
\def\){\right)}		 		 \def\BR{\Bigr)}
		 \def\BBL{\lb}
		 \def\BBR{\rb}
%
%%%%%%%%%%%%%%%%%%%%%%%%%%%%%%%%%%%%%%%%%%%%%%%%%%%%%%%%%%%%%%%
%\newcommand{\E}{{\mathbb{E}}}
%\newcommand{\1}{{\openone}}

\def\bb{{\bar{b} }}
\def\ab{{\bar{a} }}
\def\zb{{\bar{z} }}
\def\zbar{{\bar{z} }}
\def\frac#1#2{{#1 \over #2}}
\def\inv#1{{1 \over #1}}
\def\half{{1 \over 2}}
\def\d{\partial}
\def\der#1{{\partial \over \partial #1}}
\def\dd#1#2{{\partial #1 \over \partial #2}}
\def\vev#1{\langle #1 \rangle}
\def\ket#1{ | #1 \rangle}
\def\proj#1{ | #1 \rangle \langle #1 |}
\def\rvac{\hbox{$\vert 0\rangle$}}
\def\lvac{\hbox{$\langle 0 \vert $}}
\def\2pi{\hbox{$2\pi i$}}
\def\e#1{{\rm e}^{^{\textstyle #1}}}
\def\grad#1{\,\nabla\!_{{#1}}\,}
\def\dsl{\raise.15ex\hbox{/}\kern-.57em\partial}
\def\Dsl{\,\raise.15ex\hbox{/}\mkern-.13.5mu D}
\def\b#1{\mathbf{#1}}
%
%%%%%%%%%%%%%%%%%%%%GREEK LETTERS%%%%%%%%%%%%%%%%%%%%%%%%%%%%%%
%
\def\th{\theta}		 		 \def\Th{\Theta}
\def\ga{\gamma}		 		 \def\Ga{\Gamma}
\def\be{\beta}
\def\al{\alpha}
\def\ep{\epsilon}
\def\vep{\varepsilon}
\def\la{\lambda}		 \def\La{\Lambda}
\def\de{\delta}		 		 \def\De{\Delta}
\def\om{\omega}		 		 \def\Om{\Omega}
\def\sig{\sigma}		 \def\Sig{\Sigma}
\def\vphi{\varphi}
%
%%%%%%%%%%%%%%%%%%%CALIGRAPHIC LETTERS%%%%%%%%%%%%%%%%%%%%%%%%%
%
\def\CA{{\cal A}}		 \def\CB{{\cal B}}		 
\def\CC{{\cal C}}
\def\CD{{\cal D}}		 \def\CE{{\cal E}}		 
\def\CF{{\cal F}}
\def\CG{{\cal G}}		 \def\CH{{\cal H}}		 
\def\CI{{\cal J}}
\def\CJ{{\cal J}}		 \def\CK{{\cal K}}		 
\def\CL{{\cal L}}

\def\CM{{\cal M}}		 \def\CN{{\cal N}}		 
\def\CO{{\cal O}}
\def\CP{{\cal P}}		 \def\CQ{{\cal Q}}		 
\def\CR{{\cal R}}
\def\CS{{\cal S}}		 \def\CT{{\cal T}}		 
\def\CU{{\cal U}}
\def\CV{{\cal V}}		 \def\CW{{\cal W}}		 
\def\CX{{\cal X}}
\def\CY{{\cal Y}}		 \def\CZ{{\cal Z}}
\newcommand{\qed}{\rule{7pt}{7pt}}
\def\E{{\mathbf{E} }}
\def\rvac{\hbox{$\vert 0\rangle$}}
\def\lvac{\hbox{$\langle 0 \vert $}}
\def\comm#1#2{ \BBL\ #1\ ,\ #2 \BBR }
\def\2pi{\hbox{$2\pi i$}}
\def\e#1{{\rm e}^{^{\textstyle #1}}}
\def\grad#1{\,\nabla\!_{{#1}}\,}
\def\dsl{\raise.15ex\hbox{/}\kern-.57em\partial}
\def\Dsl{\,\raise.15ex\hbox{/}\mkern-.13.5mu D}
\def\beq{\begin {equation}}
\def\eeq{\end {equation}}
\def\to{\rightarrow}

\title{Distillation of local purity from quantum states}
\date{\today} 
\author{I. Devetak}
\email{devetak@csi.usc.edu}
\affiliation{IBM T.~J.~Watson Research Center, PO Box 218, Yorktown Heights, NY 10598, USA}

\begin{abstract}
Recently Horodecki et al. [Phys. Rev. Lett. 90, 100402 (2003)] introduced
an important quantum information processing paradigm,
in which two parties sharing many copies of the same bipartite quantum state 
distill local pure states, by means of  local unitary operations 
assisted by a one-way (two-way) completely dephasing channel. 
Local pure states are a valuable resource from a thermodynamical 
point of view, since they allow thermal energy to be converted into work
by local quantum heat engines.
We give a simple information-theoretical characterization of the one-way
distillable local purity, which turns out to be closely related 
to a previously known operational 
measure of classical correlations, the one-way distillable common randomness.
\end{abstract}
\maketitle

\vspace{2mm}

\section{Introduction}

One of the primary tasks of quantum information theory is
to explore the operational reductions between information processing
resources such as shared entanglement or quantum channels, 
including both the noisy and noiseless varieties. For instance, 
entanglement distillation \cite{bdsw} involves transforming a large number
of noisy bipartite quantum states $\rho^{AB}$, 
shared between two distant parties Alice and Bob, into
pure ebits $\ket{\Phi^+} = 1/\sqrt{2} (\ket{0} \ket{0} +  \ket{1} \ket{1})$
at the best possible conversion rate. This conversion task
is naturally defined within the LOCC (local operations and classical
communication) paradigm: Alice and Bob are allowed \emph{at no cost}
to i) locally add pure state ancillas to their quantum systems, 
ii) perform local unitary operations and iii) communicate classically. 
In a slight refinement of this paradigm,
one could assign a cost for one-way classical communication,
leading to trade-offs between the amount of entanglement
distilled and the classical communication invested \cite{family}. 
The communication theorist still feels at home with this
modification: after all, classical communication is a valuable 
bipartite resource, and should not be taken for granted.
It is only recently that attention has been given to \emph{local}
resources, in particular local pure states \cite{hhhhoss}. 

Local pure states can be seen as valuable from a 
\emph{thermodynamical} perspective. Although we use
the language of quantum states, the phenomenon is essentially
classical. Landauer \cite{landauer} 
was the first to observe that work was
required to erase a bit of information, i.e.
to reset a system from an unknown state to a known (pure) state.
Conversely, a supply of pure states can be used as ``fuel'' to
increase the amount of useful work extractable from a system 
at non-zero temperature \cite{szilard, bennett}. 
This is achieved by reversibly transferring entropy from the system to 
the pure states, thereby ``cooling'' the system \cite{lloyd}.

Having an appreciation for the value of pure states,
it is natural to ask about the different ways  in which they can be 
produced.
In \cite{ohhh, hhhhoss} the idea of manipulating and 
concentrating ``purity'' already existing in a diluted form,
rather than performing work to create it, was introduced.
This is very much analogous to entanglement distillation:
given a noisy resource one wishes to remove impurities from it.  
There is a local and distributed version of this problem.
In the local scenario, which we call 
\emph{purity concentration}, Alice is given a large supply of 
states $\rho^A$  and her task is to extract pure qubit states 
using only unitary operations. The maximal asymptotic rate at which 
this can be done is given by the difference between the size of the 
system $A$ (in qubits) and its von Neumann entropy \cite{hho}.
In the distributed scenario --  \emph{local purity distillation} -- 
Alice and Bob share a supply of
bipartite states $\rho^{AB}$ and they wish to distill
local pure states using CLOCC(\emph{closed} local operations
and classical communication)\cite{ohhh}, a modification of the LOCC
paradigm that disallows unrestricted consumption of local 
pure states. Horodecki et al.\cite{hhhhoss} had previously obtained some
bounds on this problem, both for the one-way and two-way
CLOCC case. 

In this paper we investigate the two scenarios 
in detail. Our main result pertains to the distributed case;
we give an information theoretical expression for the optimal one-way 
distillable local purity. This quantity turns out to be related
to a previously known operational measure of classical correlations, 
the one-way distillable common randomness \cite{cr}.
% We also show NLOCC=CLOCC for this case.
Section II is devoted to establishing notation.
Section III treats the local scenario, reproducing
the results of \cite{hho} in a somewhat more rigorous coding-theoretical 
language. The two-party distributed scenario is
considered in section IV and our main result is proved.
Section V discusses how to embed purity distillation and the CLOCC 
paradigm in the existing formalism for quantum Shannon theory,
and concludes with open questions. Appendix A collects 
a number of auxiliary inequalities used throughout the paper.
 
\section{Notation and definitions}
 
Recall the notion of an \emph{ensemble}
of quantum states $\CE = (p(x), \rho_x^{B})_{x \in \CX}$:
the quantum system $B$ is in the 
state $\rho_x^B$ with probability $p(x)$.
The ensemble $\CE$ is equivalently represented by a
\emph{classical-quantum} system \cite{cr} $X B$
in the state
\beq
\rho^{XB} = \sum_{x \in \CX} p(x) \ket{x} \bra{x}^X \otimes \rho^B_x,
\label{koki}
\eeq
where $\CH_X$ has a preferred orthonormal 
basis $\{ \ket{x} \}_{x \in \CX}$.
$X$ plays the dual role of 
an auxiliary quantum system in the
state $\sum_x p(x) \ket{x} \bra{x}$ and of 
a random variable with distribution $p$ and cardinality 
$|X|:= |\CX|$.
For a multi-party state such as $\rho^{XB}$, the
reduced density operator $\rho^{B}$ is defined by 
$\tr_{\!X} \rho^{XB}$. Conversely, we 
call $\rho^{XB}$ an \emph{extension} of $\rho^B$.
A pure extension is conventionally called a 
\emph{purification}.

The ensemble $\CE$ may come about by performing
a POVM  $\Lambda = (\Lambda_x)_x$, $\sum \Lambda_x = I$, on the $A$ part of
a bipartite state $\rho^{AB}$, in which case
$p(x) = \tr( \Lambda_x \rho^{A})$ and 
$\rho_x^{B} = p(x)^{-1} \tr_{\!A}( (\Lambda_x^A  \otimes \1^B  )\rho^{AB})$.
Equivalently, $\Lambda$ may be thought of as a quantum 
map $\Lambda: \CH_A \rightarrow \CH_X$, sending $\rho^{AB}$ to
$\rho^{XB}$. A classical map $f: \CX \rightarrow \CY$ may
similarly be viewed as a quantum one $f: \CH_X \rightarrow \CH_Y$,
$$
f(\rho) = \sum_{x \in \CX} \bra{x}\rho \ket{x} \, \ket{f(x)} \bra{f(x)}^Y, 
$$
%mapping $\rho^{XB}$ to 
%$$
%\sum_{x \in \CX} p(x) \proj{f(x)}^Y \otimes \rho^B_x,
%\label{boki}
%$$
where $\CH_Y$ has a preferred orthonormal 
basis $\{ \ket{y} \}_{y \in \CY}$.

Define the von Neumann entropy of a quantum state $\rho$
by $H(\rho) = - \tr( \rho \log \rho)$. 
We write $H(A)_\sigma = H(\sigma^A)$,
omitting the subscript when the reference state is clear from the context.
The Shannon entropy $-\sum_x p(x) \log p(x)$ 
of the random variable $X$ is thus equal to the von Neumann entropy $H(X)$
of the system $X$. 
Define the 
conditional entropy 
$$
H(A|B) = H(B) - H(AB),
$$
(quantum) mutual information 
$$
I(A;B) = H(A) + H(B) - H(AB),
$$
and conditional mutual information
$$
I(A;B|X) = I(A;BX) - I(A;X).
$$
%The \emph{coherent information} $I(A\, \rangle B)$
%is defined as $- H(A|B)$. 

For a sequence $x_1\dots x_n$ of classical indices $x_i$ 
we use the shorthand notation $x^n$, and 
$\rho_{x^n}:= \bigotimes_i \rho_{x_i}$. For
an integer $\mu$ define $[\mu] = \{1, \dots, \mu \}$.

The trace norm of an operator is defined as
$$
\|\omega \|_1 = \tr \sqrt{\omega^\dagger \omega}, 
$$ 
which for $\omega$ Hermitian amounts to the sum of the absolute 
values of the eigenvalues of $\omega$.
We say that two states $\rho$ and $\omega$ are $\epsilon$-close if
$$
\|\rho - \omega\|_1  \leq \epsilon.
$$

We  loosely refer to an isometry  $U: \CH_A  \rightarrow \CH_B \otimes \CH_C$
 as a unitary operation under the assumption that 
$A$ may be written as a composite system $BC$. For a POVM
$\Lambda = (\Lambda_x)_x$ acting on a composite system $AB$ we
say that it is \emph{rank-1 on $A$} if, for all $x$, $\Lambda_x$ is
of the form 
$$
\Lambda^{AB}_x = \proj{\phi_x}^{A} \otimes \1^B.
$$
Throughout the paper, $\ket{0}^{A}$ will denote a 
standard pure state on the system $A$.

\section{Local scenario: purity concentration}
We begin by formally defining a purity concentration code.
Alice has $n$ copies of a state $\rho^A$ defined on a system $A$ 
of dimension $d_A$.
In other words, Alice has a $n$-partite quantum system $A^n = A_1 \dots A_n$
with Hilbert space $\CH_{A^n} = \CH_{A_1} \otimes \dots 
\otimes \CH_{A_n}$ in a tensor power state $\rho^{\otimes n}$.
An $(n,\epsilon)$ \emph{purity concentration code} consists of a 
unitary operation 
$U: \CH_{A^n} \rightarrow \CH_{A_p} \otimes \CH_{A_g}$ such that, 
for $\sigma^{A_p A_g} = U(\rho^{\otimes n})$, 
\beq
\| \sigma^{A_p} - \proj{0}^{A_p} \|_1 \leq \epsilon. 
\label{uvjet}
\eeq

The \emph{rate} of the code is defined by 
$R = \frac{1}{n} \log d_{A_p}$, where $d_{A_p}$ is
shorthand for $\dim \CH_{A_p}$.
A rate $R$ is said to be \emph{achievable} if for all $\epsilon, 
\delta > 0$ and sufficiently large $n$ there exists an 
$(n, \epsilon)$ code with rate $R - \delta$.
The \emph{purity} $\kappa(\rho)$ (also referred to as ``information'' 
in \cite{hhhhoss})
is defined as the supremum over
all achievable rates $R$. 

The following theorem, previously proven in \cite{hho},
gives an information-theoretical expression for $\kappa$. 

\begin{thm}
The purity  of the state $\rho^A$
of the $d_A$ dimensional quantum system $A$ is 
$$
\kappa(\rho^{A}) = \log d_A - H(A)_\rho.
$$
\end{thm}

\vspace{2mm}

\noindent {\bf Proof} \,\, 
We start by proving the ``converse'', i.e. the $\leq$ direction 
of the theorem.
Consider a general $(n, \epsilon)$ purity concentration protocol.
 Obviously, 
$$
\log d_{A_p} = n \log d_A - \log d_{A_g}.
$$
The second term is bounded as
\begin{eqnarray}
\log d_{A_g} & \geq & H(A_g) \nonumber \\ 
& \geq & H(A_p A_g) - H(A_p) \nonumber \\
& = & n H(A) - H(A_p) \nonumber \\
& \geq & n H(A) - \frac{1}{e} - n \epsilon \log d_A.
\label{con1}
\end{eqnarray}
The second inequality follows from the subadditivity of 
von Neumann entropy (\ref{sub}), 
and the third inequality is Fannes' inequality (\ref{fannes})
applied to (\ref{uvjet}).
Hence, 
$$
R = \frac{1}{n} \log d_{A_p} \leq \log d_A - H(A) + \delta,
$$
where w.l.o.g. $\delta \geq \frac{1}{en} + \epsilon \log d_A$.

\vspace{1mm}

To prove the ``direct coding theorem'' (the $\geq$ direction),
consider the typical projector \cite{schumacher}
$\Pi^n_{\rho, \delta}$ commuting with $\rho^{\otimes n}$ with the property that, 
for all $\epsilon, \delta >0 $ and 
sufficiently large $n$ 
$$
\tr \rho^{\otimes n} \Pi^n_{\rho, \delta} \geq 1 - \epsilon,
$$  
while $\tr \Pi^n_{\rho, \delta} \leq n (H(\rho) + \delta)$. The coding theorem
now follows from lemma 1 below.
 \qed

\vspace{3mm}

\begin{lemma}
Let $\Pi$ be a projector with $\tr \Pi = d_1$ and $\rho$ a state
that commutes with $\Pi$, 
both defined on a $d_1d_2$-dimensional Hilbert space
$\CH_A$. If $\tr \rho \Pi \geq 1 - \epsilon$,
then there exists a unitary 
$U: \CH_{A} \rightarrow \CH_{B} \otimes \CH_{C}$,
with $\dim \CH_B = d_1$ and $\dim \CH_C = d_2$,
such that
$$
\| U \rho U^\dagger -  (\Pi \rho \Pi)^B  \otimes \proj{0}^C \|_1 
\leq \epsilon.
$$
\end{lemma}

\noindent {\bf Proof} \,\,\,
Let $\{ \ket{i} \}_{ i \in [d_1d_2]}$ be a basis
for $A$ such that
$$
\Pi^{A} = \sum_{i = 1}^{d_1} \proj{i}^{A}.
$$ 
Viewing $A$ as a composite system $BC$, with
basis $\{ \ket{i} \otimes \ket{j} \}_{i \in [d_1], j+1 \in [d_2]}$,
define $U$ to satisfy $U \ket{i}^A = \ket{i}^{B} \ket{0}^{C}$
for all $i \in [d_1]$. The lemma follows from
$$
\| \Pi \rho \Pi - \rho \| \leq \epsilon.
$$
%Note that for the non-commuting case the gentle
%operator lemma \cite{strong} gives an estimate of
%$\sqrt{8 \epsilon}$. 
\qed

\vspace{2mm}

\section{Bipartite scenario: local purity distillation}
We now consider the bipartite scenario where Alice and Bob share many copies
of a some state $\rho^{AB}$. Their task is to distill local pure qubit 
states by means of protocols involving only closed local operations and 
classical communication (CLOCC). More precisely,
Alice and Bob may perform local unitary operations 
and are allowed unlimited use of a completely dephasing channel in both directions. 
A dephasing channel is given by the map 
$\CP: \CH_X \rightarrow \CH_X$,  
$$
\CP(\rho) = \sum_{x} \proj{x} \rho \proj{x},
$$
where $\{ \ket{x} \}$ is an orthonormal basis for $\CH_X$.
The term ``closed'' refers to Alice and Bob
not being given free access to local pure state ancillas; this is 
the main difference between CLOCC and the more familiar LOCC relevant for
entanglement distillation \cite{bdsw}. A catalytic variation of CLOCC, 
which we denote by CLOCC$'$, allows Alice and Bob to borrow local pure state 
ancillas, but they have to return them at the end of the protocol.
Similarly define the 1-CLOCC and 1-CLOCC$'$ paradigms with the
bidirectional communication replaced by a one-way dephasing 
channel from Alice to Bob. In \cite{hhhhoss} yet another paradigm, 
NLOCC (\emph{noisy} local 
operations and classical communication) was used, which
allows both parties unlimited access to maximally mixed local states.
This additional resource will prove to be useless for our purposes. 

Our main focus will be on the 1-CLOCC$'$ paradigm as 
it turns out to be amenable to information theoretical
characterization. 
We proceed to formally define a local purity distillation code.
Alice and Bob share $n$ copies of the state $\rho^{AB}$,
embodied in the shared quantum system $A^nB^n$, and
Alice also has access to some quantum system $C$ of dimension
$d_C$, initially in a pure state $\ket{0}^C$. An $(n,\epsilon)$ 
\emph{(catalytic) 1-way local purity distillation code} consists of
\begin{itemize}
\item{a unitary operation 
$U_A: \CH_{A^n} \otimes \CH_{C} \rightarrow \CH_{A_p} \otimes \CH_{X}$ on
Alice's side}
\item a dephasing channel $\CP: \CH_{X} \rightarrow \CH_{X}$
from Alice to Bob.
\item {a unitary operation 
$U_B: \CH_{B^n}  \otimes \CH_{X} \rightarrow \CH_{B_p} \otimes \CH_{B_g}$ 
on Bob's side},
\end{itemize}
such that, for
$$
\sigma^{A_p B_p B_g} = (U_B \circ \CP \circ U_A) ((\rho^{AB})^{\otimes n} \otimes 
\proj{0}^C),
$$ 
%$$
%\sigma^{A_p B_p B_g} = 
%((\1^{A_p} \otimes U_B) \circ (\1^{A_p B^n} \otimes \CP) \circ 
%(\1^{B^n} \otimes U_A) ) 
%((\rho^{AB})^{\otimes n} \otimes 
%\proj{0}^C),
%$$ 
\beq
\| \sigma^{A_p B_p} - \proj{0}^{A_p} \otimes \proj{0}^{B_p}\|_1 \leq \epsilon. 
\label{uvjet2}
\eeq
The \emph{rate} of the code is defined by 
$R = \frac{1}{n} (\log d_{A_p B_p} - 
\log d_{C})$. The \emph{catalyst rate} 
is $\frac{1}{n} \log d_C$.
A rate $R$ is said to be \emph{achievable} if  for all $\epsilon, 
\delta > 0$ and sufficiently large $n$ there exists an 
$(n, \epsilon)$ code with rate $R - \delta$.
The \emph{1-way local purity} $\kappa_{\rightarrow}(\rho^{AB})$ is defined as 
the supremum over all achievable rates $R$. 

A quantity of particular interest is the \emph{classical deficit}
$$
\Delta^c_\rightarrow(\rho^{AB}) = 
\kappa_{\rightarrow}(\rho^{AB}) - \kappa(\rho^{A}) - \kappa(\rho^{B}).
$$
This quantity (or, rather, its bidirectional version) 
was introduced in \cite{ohhhh}, and 
advertised as a measure of classical correlations 
in the state $\rho^{AB}$.
%; indeed if the system $A$ 
%were classical it could be perfectly transmitted    
%through the dephasing channel, giving $\Delta_\rightarrow = 0$.
%\noindent {\bf Example} \,\,
\begin{eg}
Assume that Alice and Bob are given a bit of common
randomness, which is represented by the state
$$
\bar{\Phi}^{AB} = \frac{1}{2} 
(\proj{0}^A \otimes \proj{0}^B + \proj{1}^A \otimes \proj{1}^B).
$$
Alice sends her system to Bob through the dephasing channel, which
leaves it intact.
Bob performs the controlled unitary
$$
U^{AB} = \proj{0}^A \otimes \1  + \proj{1}^A \otimes V^B, 
$$
where $V \ket{1} = \ket{0}$, leaving the $B$ system
in the state $\ket{0}^B$. This gives 
$\kappa_\rightarrow = \Delta^c_\rightarrow = 1$.
\end{eg}
%\vspace{2mm}

Our main result is contained in the following theorem.
\begin{thm}
\label{theo2}
The local 1-way purity of a state $\rho^{AB}$ defined on a system of dimension 
$d_A \times d_B$ is given by 
$$
\kappa_{\rightarrow}(\rho^{AB}) = \log d_A + \log d_B - H(A)_\rho - H(B)_\rho +
D_{\rightarrow}(\rho^{AB}),
$$
with 
$$
D_\rightarrow(\rho^{AB}) = \lim_{n \rightarrow \infty}
\frac{1}{n} D^{(1)}_\rightarrow  \left( (\rho^{AB})^{\otimes n}\right)
$$
and 
\beq
D^{(1)}_\rightarrow(\rho^{AB}) = \max_{\Lambda}
I(X;B)_{(\Lambda \otimes \1)(\rho)}.
\label{dager}
\eeq
The maximization is over all rank-1 POVMs $\Lambda:\CH_A \rightarrow \CH_X$. 
\end{thm}
\begin{cor}
$$
\Delta^c_\rightarrow(\rho^{AB}) =  D_{\rightarrow}(\rho^{AB}).
$$
\end{cor}

\vspace{2mm}
The quantity $D^{(1)}_\rightarrow(\rho^{AB})$ first 
appeared in \cite{hv}, where it was proposed, on heuristic grounds,
as a measure of classical correlations in the state $\rho^{AB}$. 
Its ``regularized'' version $D_\rightarrow(\rho^{AB})$ \cite{ohhhh}
was given operational meaning in \cite{cr} where 
it was shown to be equal to the \emph{1-way distillable common randomness}
(1-DCR) of $\rho^{AB}$. The 1-DCR is the
maximum conversion rate from $\rho^{AB}$ 
into  bits of common randomness,
achievable with 1-LOCC, \emph{in excess} of the classical communication invested. 

In \cite{cr}, the  additivity of $D^{(1)}_\rightarrow$
was shown for a separable state $\sigma^{AB}$
and arbitrary $\rho^{AB}$,
\beq
D^{(1)}_\rightarrow(\rho^{AB} \otimes \sigma^{AB}) 
= 
D^{(1)}_\rightarrow(\rho^{AB}) + D^{(1)}_\rightarrow(\sigma^{AB}). 
\label{trog}
\eeq
Therefore, adding local maximally mixed states 
$\sigma^{AB} = (d_A d_B)^{-1} \, \1^{A} \otimes \1^{B}$,
for which $D^{(1)}_\rightarrow(\sigma^{AB}) = 0$
does not affect the 1-DCR or the classical deficit.
Moreover, for separable states $\rho^{AB}$ the 
classical deficit is efficiently computable, as
$$
D_\rightarrow(\rho^{AB}) =  D^{(1)}_\rightarrow(\rho^{AB}).
$$
%Note that for separable states that are not
%of the classical-quantum form (\ref{koki}), the
%work deficit is positive. This, unfortunately, 
%is an argument \emph{against} viewing the
%work deficit as a measure of quantum correlations.
From \cite{cr} we know additivity to hold for 
the case of pure states $\ket{\phi}^{AB}$, 
and it is easily seen that \cite{hhhhoss}
$$
\Delta^c_\rightarrow(\phi^{AB}) = E(\phi^{AB}):= 
H(A)_\phi,
$$
where $E$ is the unique measure of entanglement for pure states.
Additivity also holds for Bell-diagonal states \cite{thld, hhosss}.
The general question of the additivity of $D^{(1)}_\rightarrow$
is known to be equivalent to several other open additivity
problems in quantum information theory \cite{kw, shor, msw}, 
including that of the Holevo capacity of quantum channels.

In proving theorem \ref{theo2}, 
we shall need two lemmas. The first is from \cite{sw}:
\begin{lemma}
\label{dva}
Consider a classical-quantum system $X^n B^n$
in the state $(\rho^{XB})^{\otimes n}$, where 
$\rho^{XB}$ is given by (\ref{koki}).
For any $\epsilon, \delta > 0$ and sufficiently large $n$, 
there exist 
\begin{itemize}
\item a set $\CS$ in $\CX^n$ with 
\beq
\Pr \{X^n \notin  \CS \} \leq \epsilon,
\label{xstuff}
\eeq
\item a bijection $f: [\mu] \times [\lambda]  \rightarrow \CS $, where 
$\lambda \leq 2^{n [H(X) - I(X;B) + \delta]}$ and
$\mu \lambda \leq 2^{n [H(X) + \delta]}$,
\item a collection of POVMs
$(\Upsilon^{(l)})_{l \in [\lambda]}$
(each  $\Upsilon^{(l)} = (\Upsilon^{(l)}_m)_m$ is a POVM),
% $\sum_m \Upsilon^{(l)}_m = I, \,\,\, 
%\forall l \in [\lambda]$, 
such that 
\beq
\tr \rho^B_{f(m,l)} \Upsilon^{(l)}_m \geq 1 - \epsilon, \,\,\, \forall m,l. 
\label{msuc}
\eeq
\end{itemize}
\end{lemma}
The above lemma says that a highly probable set $\CS$ of sequences 
$x^n$ can be covered by $\lambda$ disjoint sets $\CS_l$, $l \in [\lambda]$,
of size $\mu$ in such a way that, given the index $l$,  the identity
of a particular  sequence  in $\CS_l$ may be reliably inferred from a measurement
on $B^n$. 

The following technical lemma is a corollary of the measurement compression
theorem \cite{winter}, and is proved in appendix B.
\begin{lemma}
\label{aw}
Given the system $A^n B^n$ in the state $(\rho^{AB})^{\otimes n}$
and a rank-1 POVM $\Lambda: \CH_A \rightarrow \CH_X$,
for any $\epsilon, \delta > 0$
and sufficiently large $n$, there exists 
\begin{itemize}
\item a decomposition $A^n = A_1 A_2$ such that
$$
H(A_1) \leq n \epsilon
$$
\item a POVM $\tilde{\Lambda}: \CH_{A^n}  \rightarrow \CH_K$
which is rank-1 on $A_2$ and
\begin{eqnarray}
%\log |K|
\log |K| & \leq  & n[H(A) + \delta] \label{uno} \\
I(K; B^n)_\omega & \geq & n [I(X; B)_\rho - \epsilon] \label{due},
\end{eqnarray}
where 
\begin{eqnarray}
\rho^{X B} & =  & ({\Lambda} \otimes  \1) (\rho^{AB}),
\label{lemro}\\
\omega^{K B^n} & =  & (\tilde{\Lambda} \otimes  \1) (\rho^{AB})^{\otimes n}.
\end{eqnarray}
\end{itemize}
\end{lemma}

\vspace{2mm}

\noindent {\bf Proof of theorem 2} \,\, First, let us prove the converse.
Consider a general $(n, \epsilon)$ purity distillation protocol.
We know that 
$$
\log d_{A_p B_p} - \log d_C =  n (\log d_A + \log d_B)
 - \log d_{B_g}.
$$
Assume, w.l.o.g., $\delta \geq \frac{1}{en} + \epsilon \log (d_A d_B)$.
The entropic quantities below refer 
to the overall quantum state at a stage of the protocol
which is implicit from the subsystems involved.
For instance, the system $B^n$ exists only before $U_B$ is applied.
\begin{eqnarray*}
\log d_{B_g} & \geq & H(B_g) \\
& \geq & H(B_p B_g) - H(B_p) \\
& =  & H(X B^n) - H(B_p) \\
& \geq & H(X) + H(B^n|X) - \frac{1}{e} - n \epsilon \log d_A \\
& \geq & H(A_g) + H(B^n|X) - \frac{1}{e} - n \epsilon \log d_A \\
& \geq & n H(A) + H(B^n|X) - n \delta.
\end{eqnarray*}
The second inequality is subadditivity (\ref{sub}) , the
third is Fannes' inequality (\ref{fannes}) and (\ref{uvjet2}), 
the fourth follows from the fact that
dephasing cannot decrease entropy \cite{nie&chuang} and the fifth
follows  along the lines of (\ref{con1}).
Hence, 
\begin{eqnarray*}
R & = & \frac{1}{n} (\log d_{A_pB_p} - \log d_C) \\
& \leq &
\log d_A + \log d_B  - H(A) - H(B) +
\frac{1}{n} I(X; B^n) +    \delta.
\end{eqnarray*}

The idea behind the direct coding theorem is
that there are two potential sources
of purity. The first comprises the locally concentrable
purity for the two parties, from section III, and is responsible
for the $\kappa(\rho^A) + \kappa(\rho^B)$ term.
The second comes from the classical correlations
present in the system, and gives rise to the
$D_\rightarrow(\rho^{AB})$ term. Roughly speaking, Alice sends her part
of the classical correlations through the
dephasing channel; Bob then takes advantage of
the redundancy, as in example 1, to distill purity.

We start by considering a special case. 
Assume that the system
$A$ can be divided into subsystems $A = A_1 A_2$
such that $H(A_1) \leq \tau$,
and that $\Lambda$ is rank-1 on $A_2$.
We show that we can achieve a rate arbitrarily close to
$$
\log d_A + \log d_B - \tau - H(X)_\rho - H(B)_\rho +
I(X;B)_\rho,
$$
with $\rho$ given by (\ref{lemro}).
Consider a sufficiently large $n$ and the induced decomposition
$A^n = A_1^n A_2^n$.
The purity distillation protocol comprises of the following steps.
\begin{enumerate}
\item First, Alice applies the protocol from theorem 1 to
$A_1^n$, yielding a subsystem $A_{1p}$ of size 
$n [\log d_{A_1} - \tau  - \delta]$ qubits, 
in a state $\epsilon$-close to $\ket{0}^{A_{1p}}$.
\item The measurement $\Lambda^{\otimes n}$ may be 
implemented by borrowing $n \log d_X$ qubit ancillas 
(in some fixed state $\ket{0}^{X^n}$), 
performing some unitary operation $U$ on the system 
$A_2^n X^n$, and completely dephasing the system
$X^n$ in a fixed basis $\{ \ket{x^n} \}$.
Here we let Alice perform this measurement \emph{coherently},
i.e. by omitting the dephasing step (the channel $\CP$ will 
later do this for us). 
Since $\Lambda^{\otimes n}$ is rank-1 on $A_2^n$, this results in 
a state of the form
$$
\sum_{x^n} \sqrt{p(x^n)} \ket{x^n}^{X^n} \ket{\psi_{x^n}}^{A_2^n}
\ket{\phi_{x^n}}^{R^n}, 
$$ 
where $R^n$ is the ``reference system'' that purifies the initial state
of $A_2^n$.
She then performs the controlled unitary
$$
\sum_{x^n} \proj{x^n}^{X^n} \otimes  V_{x^n}^{A_2^n},
$$
where $V_{x^n} \ket{\psi_{x^n}} = \ket{0}$,
leaving $A_2^n$ in the state $\ket{0}^{A_2^n}$.
% Composing
%the two unitaries, we see that a catalyst of only 
%$n [\log d_{X} - \log d_{A_2}]$ qubits would have sufficed.
\item Were Alice to perform the von Neumann measurement
on $X^n$, the resulting state of the system $X^n B^n$ would
be 
$$
(\rho^{XB})^{\otimes n} =
\sum_{x^n} p(x^n) \proj{x^n}^{X^n} \otimes 
\rho_{x^n}^{B^n}.
$$
%Define $\theta^{X^n}$ to be the ``dephased'' 
%state of the system $X^n$, i.e., with the 
%off-diagonal parts (in the $\{ \ket{x^n} \}$ basis) removed. 
Choose the set $\CS$, bijection $f$ and 
collection of POVMs $(\Upsilon^{(l)})_l$ as in lemma \ref{dva}.
Define $\Pi' = \sum_{x^n \in \CS} \proj{x^n}^{X^n} \otimes I^{B^n}$. 
By (\ref{xstuff}) and the proof of lemma 1, there is a unitary operation 
(acting on Alice's system only!) that takes $(\rho^{XB})^{\otimes n}$
to a state $2 \epsilon$-close to 
$\proj{0}^{X'} \otimes {\theta'}^{M L B^n}$ with  
$d_{X'} = (\mu \lambda)^{-1} {d_X}^{\!\! n}$ and
$$
{\theta'}^{M L B^n} = 
\sum_{m,l} p(m,l) \proj{m}^{M} \otimes \proj{l}^{L}
\otimes \rho_{f(m,l)}^{B^n}.
$$
The $p(m,l)$ is some probability distribution associated
with a composite random variable $ML$. Alice performs
said unitary.
\item Alice sends the $ML$ system through the dephasing channel, leaving
$MLB^n$ in a state $\theta^{MLB^n}$ which is
$2 \epsilon$-close to ${\theta'}^{MLB^n}$.
\item For each $l$ one can define
a unitary $W_l^{B^n M}$, a coherent version of the measurement $\Upsilon^{(l)}$,
which upon measurement ``outcome'' $m$ performs
the transformation $\ket{m}^M \mapsto  \ket{0}^M$.
Explicitly, $W_l^{B^n M}$ is w.l.o.g. of the form 
$\sum_{m,m'} \ket{m'}\bra{m}^M \otimes Y_{m'm}^{B^n}$.
Choosing $Y_{0m} = (\Upsilon^{(l)}_m)^{\frac{1}{2}}$ and the remaining
$Y_{m'm}$ to satisfy unitarity leaves $W_l$ with the desired 
property. Defining
$$
\sigma^{B^n M}_{ml} = 
W_l^{B^n M} (\rho^{B^n}_{f(m,l)} \otimes \proj{m}^{{M}}),
$$
the measurement success criterion (\ref{msuc}) of lemma \ref{dva}
becomes 
$$
\bra{0} \sigma^{M}_{ml} \ket{0} \geq 1 - \epsilon.
$$ 
By (\ref{rut}),
\beq
\| \sum_{m,l} p(m,l) \sigma^{M}_{ml} - \proj{0}^M \|_1 \leq 2 \sqrt{\epsilon}.
\label{sitar}
\eeq
%Then $W_l^{-1}$ resets $m$ to $0$.
Bob  applies the controlled unitary
$$
W^{L B^n M} = \sum_{l} \proj{l}^L \otimes W_l^{B^n M},
$$
which, by (\ref{sitar}), maps ${\theta'}^{M L B^n}$ to a state whose $M$ 
part is $2 \sqrt{\epsilon}$-close to $\ket{0}^M$. 
Since $\theta^{M L B^n}$ is $2\epsilon$-close
to ${\theta'}^{M L B^n}$, upon application of $W$ its
$M$ part  becomes 
$( 2 \epsilon + 2 \sqrt{\epsilon} )$-close to
$\ket{0}^M$, by the triangle inequality (\ref{tri}). 
\item By the gentle operator lemma (see appendix A),
performing $W$ perturbs the $B$ system very little,
leaving it in a state 
$(\epsilon + \sqrt{8 \epsilon})$-close 
to $(\rho^B)^{\otimes n}$.
Bob applies the protocol from theorem 1 to $B^n$,
yielding a subsystem $B_{p}$ of size $n (d_B - H(B) - \delta)$
qubits, in a state $(2 \epsilon +  \sqrt{8 \epsilon})$-close 
to $\ket{0}^{B_p}$.
\end{enumerate}

In summary, the protocol consumes a catalyst of 
$n \log d_X $ qubits, while returning
a system of size
\begin{eqnarray*}
 & n [\log d_{A_1} - \tau - \delta] +  n \log d_{A_2} + 
n \log d_X - \log (\mu \lambda) \\
 &\phantom{==}  +  \log {\mu} +  n [d_{B} - H(B) - \delta]
\end{eqnarray*}
qubits, in a state which is 
$(7 \epsilon + (2 + \sqrt{8}) \sqrt{\epsilon}) $-close to pure.
This corresponds to a purity distillation rate of at least
$$ 
\log d_A + \log d_B - \tau - H(X) - H(B) + I(X;B) - 3 \delta,
$$ 
while the classical communication rate required was 
$n^{-1}\log (\mu \lambda) \leq H(X)+ \delta$ bits per
copy.

To prove the general statement of the theorem we shall
rely on lemma \ref{aw}  and ``double blocking''.
Let $n'$ be  sufficiently large for lemma \ref{aw} to apply
with respect to the \emph{optimal}
$\Lambda$ achieving $D^{(1)}_\rightarrow(\rho^{AB})$ in (\ref{dager}). 
We shall apply the special-case protocol described above
to the block system $A^{n'} = A_1 A_2$ and block measurement 
$\tilde{\Lambda}$, obtaining a rate of 
\begin{eqnarray*}
&\log d_A + \log d_B - 
\frac{1}{n'} H(K) - H(B)
% \\ 
%& \phantom{====} 
+  \frac{1}{n'} I(K; B^n) - 3 \delta - \epsilon.
\end{eqnarray*}
%with respect to the source $(\rho^{AB})^{\otimes n' n}$.
By lemma \ref{aw} and (\ref{dager}), this is bounded from below by
\begin{eqnarray*}
&\log d_A + \log d_B - 
 H(A)- H(B) 
%\\ 
%& \phantom{==}  
+  D^{(1)}_{\rightarrow}(\rho^{AB}) - 4 \delta - 2\epsilon.
\end{eqnarray*}
%Note that the rate at which the catalyst is needed 
%is bounded from above by $\delta$, i.e. it can be made
%arbitrarily small.
The classical communication rate required for this protocol
is  $H(A) + 2 \delta$. 

Finally, a third layer of blocking allows us to replace 
$D^{(1)}_{\rightarrow}$ by $D_{\rightarrow}$,
and we are done. \qed 
\vspace{6mm}

%Or: ORGANIZE by saying ``compressing doesn't disturb the system very much,
%in the sense of entanglement''. 
%Do this in theorem 1.

\vspace{3mm}
 
It is not hard to see that the above protocol
may be bootstrapped to make the catalyst rate
arbitrarily small. Moreover, if $\kappa(\rho^A) > 0$
a catalyst is not needed at all (see also \cite{hhosss}). 

\section{Discussion}

The question of counting local resources in 
standard quantum information theoretical tasks, 
such as entanglement distillation, was
recently raised by Bennett \cite{charlie}.
In particular, it is desirable to 
extend the theory of \emph{resource inequalities}
\cite{family} to include the manipulation of local resources. 
Recall the notation from \cite{cr} in which 
$[c \rightarrow c]$, $[q \rightarrow q]$ and $[q  q]$
stand for  a bit of classical communication, a qubit of quantum
communication and and ebit of entanglement, respectively.
There it was implicit that local pure
ancillas could be added for free, which makes a classical channel
and a dephasing quantum channel operationally equivalent. 
To define  a ``closed'' version of this formalism, one
must identify $[c \rightarrow c]$ with a dephasing qubit channel,
and introduce a new 
resource: a \emph{pbit} of purity, defined as a local pure qubit state $\ket{0}$
w.l.o.g. in Bob's posession.
A pbit may be written
as either $[q]$ or $[c]$, as there is little distinction between classical
and quantum for strictly local resources. 
The main result of our paper may be written succinctly as
$$
\{ qq \} + H(A)_\rho \,[c \rightarrow c]  \geq  
\kappa_{\rightarrow}(\rho^{AB}) \, [q],
$$
where  $\{ q q \}$ represents the noisy static resource $\rho^{AB}$,
and $\kappa_{\rightarrow}(\rho^{AB})$ is given by theorem 2.
Regarding entanglement distillation, 
closer inspection of the optimal one-way 
protocol from \cite{devetak:winter}
reveals that 
\begin{itemize}
\item only a negligible rate of pure state 
ancillas need be  consumed
\item  moreover, the locally concentrable purity 
$\kappa(\rho^A) + \kappa(\rho^B)$ is available
without affecting the entanglement distillation rate.
\end{itemize}
Whether the above holds for general quantum Shannon theoretic problems
remains to be investigated. 

We conclude with a list of open problems. 
\begin{enumerate}
\item
It would
be interesting to find the optimal trade-off
between the local purity distilled and the one-way classical
communication (dephasing) invested.
In particular, does the problem  reduce to the 1-DCR trade-off
curve from \cite{cr}? Also, one could consider 
purity distillation assited by quantum communication 
\cite{phase}.
\item We have seen that purity distillation and common randomness
distillation are intimately related. Is there a non-trivial trade-off
between the two, or it is always optimal to (linearly) 
interpolate between the known purity distillation and common
randomness distillation protocols?
One could also consider the simultaneous distillation
of purity and other resources, such as entanglement 
(see \cite{ohhhh}).  
\item Clearly, one would like a formula for the two-way
distillable local purity. Solving this problem in
the sense of the present paper appears to be difficult;
\cite{hhosss} gives a formula involving 
maximizations over a class of states which is, alas, 
rather hard to characterize.
A more tractable question is whether the relationship established
between distillable purity and distillable common randomness
carries over to the two-way scenario. 
 \end{enumerate}
\acknowledgments
We are grateful to Charles Bennett, Guido Burkard, David DiVincenzo, 
Aram Harrow, Barbara Terhal and John Smolin for useful
discussions. We also thank Micha{\l} and Pawe{\l} Horodecki, 
Jonathan Oppenheim and Barbara Synak for comments on the
manuscript and sharing their 
unpublished results on purity distillation \cite{hhosss,synak}.
This work was supported in part by the NSA under the US Army Research 
Office (ARO), 
grant numbers DAAG55-98-C-0041 and DAAD19-01-1-06.

\appendix

\section{Miscellaneous  inequalities}
For states $\rho$, $\omega$ and $\sigma$, 
the triangle inequality holds:
\beq
\|\rho - \omega \|_1 + \|\omega - \sigma \|_1
\geq \|\rho - \sigma \|_1. 
\label{tri}
\eeq
The following bound \cite{fuchs}
relates trace distance and fidelity:
\beq
\| \rho- \proj{\phi} \|_1 \leq 2 \sqrt{1 - \bra{\phi} \rho \ket{\phi}}.
\label{rut}
\eeq
The gentle operator lemma \cite{strong} says that a POVM element that 
succeeds on a state with high probability does not disturb it much.
\begin{lemma}
  \label{tender}
  For a state $\rho$ and  operator $0\leq \Lambda\leq\1$,
  if $\tr(\rho \Lambda)\geq 1-\lambda$, then
  $$\left\|\rho-\sqrt{\Lambda}\rho\sqrt{\Lambda}\right\|_1\leq \sqrt{8\lambda}.$$
  The same holds if $\rho$ is only a subnormalized
  density operator.
  \qed
\end{lemma}
For two states $\rho$ and $\omega$ defined on 
a $d$-dimensional Hilbert space, 
Fannes' inequality \cite{fannes} reads:
\beq
| H(\rho) - H(\sigma) | \leq \frac{1}{e}
 + \log d \|\rho - \omega\|_1.
\label{fannes}
\eeq
An important property
of von Neumann entropy is \emph{subadditivity}
\beq
H(B) \geq H(AB) - H(A).
\label{sub}
\eeq

\section{Proof of lemma \ref{aw}}
By the proof of the measurement compression theorem \cite{winter},
for any $\epsilon, \delta > 0$ and sufficiently large $n$ 
there is an ensemble of rank-1 sub-POVMs 
$(p_s, \tilde{\Lambda}^{(s)}: \CH_{A^n} \rightarrow \CH_K)_s$ 
and a classical map $g: \CH_S \otimes \CH_K \rightarrow 
\CH_{X^n}$ such that
\begin{itemize}
\item 
$\sum_k \tilde{\Lambda}^{(s)}_k \leq \Pi$, where
the index $k$ ranges over $[2^{n[H(A) + \delta]}]$, and
$\Pi$ is a  projector commuting with $(\rho^{A})^{\otimes n}$
such that $\tr \Pi \leq 2^{n[H(A) + \delta]}$ and 
$\tr (\rho^{A})^{\otimes n} \Pi \geq 1 - \epsilon$.
\item 
\beq
\left\| (\rho^{XB})^{\otimes n} - \sigma^{X^n B^n}  \right\|_1 \leq \epsilon,
\label{bogy}
\eeq
where 
\begin{eqnarray*}
\sigma^{X^nB^n} & =  & (g \otimes \1^{B^n}) \Omega^{S K B^n}\\ 
\Omega^{S K B^n} & = & \sum_s p(s) \proj{s}^S \otimes
 [(\Lambda^{(s)} \otimes \1^{B^n})(\rho^{AB})^{\otimes n}],
\end{eqnarray*}
for some probability distribution $p(s)$.
\end{itemize}
Each sub-POVM $\tilde{\Lambda}^{(s)}$ may be augmented by
no more than  $2^{n[H(A) + \delta]}$ rank-1 elements
to satisfy equality 
$$\sum_k \tilde{\Lambda}^{(s)}_k = \Pi.$$
The proof of lemma 1 and 
Fannes' inequality  (\ref{fannes}) imply the existence of  a decomposition
$A^{n} = A_1 A_2$ such that 
$$
H(A_1) \leq \frac{1}{e} +  n \epsilon \log d_A,
$$
while $\tilde{\Lambda}^{(s)}$ is now viewed as a rank-1 POVM 
on $A_2$ such that (\ref{bogy}) still holds for the $\tilde{\Lambda}^{(s)}$. 
% Condition (\ref{uno}) follows.

Define $\epsilon' =  \frac{3}{ne} + 2 \epsilon \log (d_X d_B)$.
Then
\begin{eqnarray*}
n I(X; B)_\rho & \leq & I(X^n; B^n)_\sigma - n \epsilon' \\
& \leq & I(KS; B^n)_\Omega -  n \epsilon' \\
& = & I(S; B^n)_\Omega + I(K; B^n| S)_\Omega -  n \epsilon' \\
& = & I(K; B^n| S)_\Omega -  n \epsilon'.
\end{eqnarray*}
The first inequality is a triple application of  Fannes' inequality,
and the second is by the data processing inequality (see e.g. \cite{nie&chuang}).
The last line is by locality: the state
of $B^n$ is independent of which measurement $\tilde{\Lambda}^{(s)}$
gets applied to $A^n$ .
Thus there exists a particular $s$ such that (\ref{due})
is satisfied for 
$\tilde{\Lambda} = \tilde{\Lambda}^{(s)}$. \qed

\end{document}